 \documentclass[final,5p,times,twocolumn]{elsarticle}

\usepackage{amssymb}
\usepackage{color}

\journal{Physics Letters B}

\begin{document}

\begin{frontmatter}



\title{Isomerism in the ``south-east'' of $^{132}$Sn and a predicted neutron-decaying isomer in $^{129}$Pd}


\author[IFCEN]{Cenxi Yuan\corref{cor1}}
\author[IMP]{Zhong Liu\corref{cor2}}
\author[PKU,CTNP]{Furong Xu}
\author[Surrey]{P. M. Walker}
\author[Surrey]{Zs. Podoly\'{a}k}
\author[NJU]{C. Xu}
\author[NJU]{Z. Z. Ren}
\author[IMP]{B. Ding}
\author[IMP]{M. L. Liu}
\author[IMP]{X.Y. Liu}
\author[IMP]{H.S. Xu }
\author[IMP]{Y.H. Zhang}
\author[IMP]{X.H. Zhou}
\author[IMP]{W. Zuo}

\address[IFCEN]{Sino-French Institute of Nuclear Engineering and Technology, Sun Yat-Sen University, Zhuhai, 519082, Guangdong, China}
\address[IMP]{Institute of Modern Physics, Chinese Academy of Sciences, Lanzhou 730000, China}
\address[PKU]{State Key Laboratory of Nuclear Physics and Technology, School of Physics, Peking University, Beijing 100871, China}
\address[CTNP]{Center for Theoretical Nuclear Physics, National Laboratory for Heavy Ion Physics, Lanzhou 730000, China}
\address[Surrey]{Department of Physics, University of Surrey, Guildford, Surrey GU2 7XH, United Kingdom}
\address[NJU]{Department of Physics, Nanjing University, Nanjing 210093, China}

\cortext[cor1]{yuancx@mail.sysu.edu.cn}
\cortext[cor2]{liuzhong@impcas.ac.cn}

\begin{abstract}
Excited states in neutron-rich nuclei located south-east of $^{132}$Sn are investigated by shell-model calculations. A new shell-model Hamiltonian is constructed for the present study. The proton-proton and neutron-neutron interactions of the Hamiltonian are obtained through the existing CD-Bonn $G$ matrix results, while the proton-neutron interaction across two major shells is derived from the monopole based universal interaction plus the M3Y spin-orbit force. The present Hamiltonian can reproduce well the experimental data available in this region, including one-neutron separation energies, level energies and the experimental $B(E2)$ values of isomers in $^{134,136,138}$Sn, $^{130}$Cd, and $^{128}$Pd. New isomers are predicted in this region, $e.g.$ in $^{135}$Sn, $^{131}$Cd, $^{129}$Pd, $^{132,134}$In and $^{130}$Ag, in which almost no excited states are known experimentally yet. In the odd-odd $^{132,134}$In and $^{130}$Ag, the predicted very long $E2$ life-times of the low-lying $5^{-}$ states are discussed, demanding more information on the related proton-neutron interaction. The low-lying states of $^{132}$In are discussed in connection with the recently observed $\gamma$ rays. The predicted $19/2^{-}$ isomer in $^{129}$Pd could decay by both electromagnetic transitions and neutron emission with comparable partial life-times, making it a good candidate for neutron radioactivity, a decay mode which is yet to be discovered.
\end{abstract}

\begin{keyword}



\end{keyword}

\end{frontmatter}


\section{\label{sec:level1}Introduction}

On the journey towards the neutron drip-line one needs a reliable theoretical model which incorporates the known features of the nuclear many-body system and has enough predictive power for a range of unexplored nuclei. The nuclear shell model is one such, providing the basic framework for understanding the detailed structure of complex nuclei as arising from the individual motion of nucleons and the effective nuclear interactions between them. In the shell model, doubly magic nuclei, especially those far from the line of stability, such as $^{132}$Sn, act as cornerstones for exploring the unknown regions.

Experimentally, the observation of isomers has been key to the understanding of the shell structure and the development of the shell model~\cite{segre1949}. Recently, nuclei around $^{132}$Sn have been the subject of intensive experimental studies with respect to the persistence of the $N = 82$ shell gap and its relevance to the astrophysical $r$-process path. Early $\beta$-decay results seemed to indicate a substantial shell quenching~\cite{dillmann2003}, while isomeric spectroscopy studies gave evidence for the persistence of the $N = 82$ shell down to $Z=46$, $^{128}$Pd~\cite{jungclaus2007,gorska2009,watanabe2013}. Along the $Z=50$ line isomeric states were also observed in $^{134,136,138}$Sn~\cite{korgul2000,simpson2014}, and mixing between seniority-2 and -4 configurations was revealed for the $6^{+}$ isomer of $^{136}$Sn~\cite{simpson2014}. Furthermore, mass measurements have been crucial for experimental determination of the shell gaps \cite{At15,Kn16}.

In addition, isomers in the region far from the $\beta$-stability line could serve as stepping stones towards the drip-lines. For example, the high-spin $(19/2^{-})$ isomer in $^{53}$Co provided the first example of proton radioactivity~\cite{jackson1970}. Similarly in the very neutron-rich region, a neutron may ``drip'' from an isomer before the neutron drip-line itself is reached, if the isomer's excitation energy takes it above the neutron separation energy~\cite{Pe71,walker2006}.

In this paper, shell-model calculations are performed to investigate the isomerism in the south-east quadrant of $^{132}$Sn, i.e. with $Z\leq 50$ and $N\geq 82$, including the possibility of neutron radioactivity from such isomers.

\section{\label{sec:level2}Effective Hamiltonian}
The construction of an effective Hamiltonian is one of the key elements in a shell-model study. The model space for the present work is $\pi0f_{5/2}$, $\pi1p_{3/2}$, $\pi1p_{1/2}$, $\pi0g_{9/2}$, and $\nu1f_{7/2}$, $\nu2p_{3/2}$, $\nu2p_{1/2}$, $\nu0h_{9/2}$, $\nu1f_{5/2}$, $\nu0i_{13/2}$, corresponding to the $Z=28-50$ and $N=82-126$ major shells, respectively. Below $^{132}$Sn, the robustness of the $N=82$ shell gap has been experimentally examined and confirmed down to Cd and Pd~\cite{jungclaus2007,watanabe2013,At15,Kn16}. The core excited states in $^{131}$In are found to be at nearly 4 MeV~\cite{gorska2009}, thus this model space is suitable for the investigation of the low-lying states around or lower than 2 MeV in Sn, In, Cd, Ag, and Pd isotopes with $N>82$. So far there is no well established effective Hamiltonian for this model space due to the lack of experimental data on the excited states in this most neutron-rich region around $^{132}$Sn. An effective Hamiltonian for this model space is proposed below. In a very recent work, shell-model calculations for $^{132}$In were performed employing a modern realistic effective interaction and two-body matrix elements deduced from the $^{208}$Pb region \cite{Jungclaus2016}.

The single-particle energies for the four proton orbits and the six neutron orbits in the present model space are fitted to the reported energies of the single-particle states of $^{131}$In and $^{133}$Sn, respectively. These energies are from Ref.~\cite{nndc} and the recently discovered $\pi1p_{1/2}$ and $\pi1p_{3/2}$ single-hole states in $^{131}$In~\cite{Kankainen2013,taprogge2014}. The single-particle energy for the $\nu0i_{13/2}$ orbit in Ref.~\cite{nndc} is estimated from the excitation energy of the $10^{+}$ state in $^{134}$Sb ~\cite{Urban1999}. The present Hamiltonian fixes the relative single-particle energies to the observed excited states. It is reasonable as the present work concentrates on the excitation energies of levels and neutron separation energies.


The proton-proton interaction is based on the proton-proton part of jj45pna Hamiltonian, which has been derived from the CD-Bonn potential through the $G$ matrix renormalization method by Hjorth-Jensen and is included in the OXBASH package~\cite{OXBASH}. The theoretical method to derive the jj45pna effective interaction and its application in the $A\sim100$ region is described in Ref. \cite{Jensen1995}. Recently, the Hamiltonian jj45pna was also used to investigate the $\beta$ decay of $^{113}$Cd and $^{115}$In~\cite{Haaranen2016}, and the low-lying states of In isotopes around A=125~\cite{Rejmund2016}. The strength of this interaction is modified by a factor $0.74$ to reproduce the low-lying states of $^{130}$Cd. The neutron-neutron interaction is from the neutron-neutron part of CWG Hamiltonian, which is derived from the CD-Bonn renormalized $G$ matrix and used to study nuclei around $^{132}$Sn~\cite{brown2005}.

The proton-neutron interaction is calculated through an effective nucleon-nucleon, monopole-based universal interaction V$_{MU}$~\cite{otsuka2010} plus a spin-orbit force from M3Y~\cite{m3y1977}(V$_{MU}$+LS). The validity of the V$_{MU}$+LS interaction in shell-model calculations has been examined in different regions of the chart of nuclei. The structure features of neutron rich C, N, O~\cite{yuan2012}, Si, S, Ar, Ca~\cite{utsuno2012}, Cr, and Fe isotopes~\cite{Togashi2015} have been nicely described by shell-model calculations by taking V$_{MU}$+LS as the proton-neutron interaction between the $p$ proton shell and the $sd$ neutron shell~\cite{yuan2012}, the $sd$ proton shell and the $pf$ neutron shell~\cite{utsuno2012}, and the $pf$ proton shell and  the $gds$ neutron shell~\cite{Togashi2015}, respectively. For example, the neutron drip-lines for carbon, nitrogen and oxygen isotopes are simultaneously explained, revealing the impact of the proton-neutron interaction on the evolution of the nuclear shell~\cite{yuan2012}. Recently, the first $4^{+}$ state of $^{44}$S has been identified as a high-$K$ isomer~\cite{Utsuno2015} through the Hamiltonian suggested in Ref.~\cite{utsuno2012}. In the heavier region, close to $^{132}$Sn, the change of the energy difference between the $10^{+}$ and $7^{-}$ yrast levels in the $N=80$ isotones down to $^{126}$Pd is well explained by V$_{MU}$+LS~\cite{Watanabe2014}. Thus it is natural and reasonable to use this interaction as the proton-neutron interaction in the present study.

In the present Hamiltonian the strength of the central-force parameters of V$_{MU}$ is enhanced $1.07$ times the original one in Ref.~\cite{otsuka2010} to give a better description of the one-neutron separation energies $S_{n}$. It should be noted that the original form of V$_{MU}$ comes from the effective Hamiltonian in the $sd$ and $pf$ regions. In $psd$ region, the strength of its central part is reduced by a factor of $0.85$ to reproduce the binding energies of the B, C, N, and O isotopes. The new Hamiltonian is named as $jj46$ in the following discussion as it includes 4 and 6 valence proton and neutron orbits, respectively. The present shell-model calculations are performed using the code OXBASH~\cite{OXBASH}.

It should be mentioned that the present Hamiltonian operates in the particle-particle model space. The doubly magic nucleus $^{132}$Sn has fully occupied valence protons and no valence neutrons in the present model space. Starting from $^{132}$Sn, the proton-hole energies of $^{131}$In and the neutron-particle energies of $^{133}$Sn are not directly taken as the single particle energies in the Hamiltonian, but are modified by the residual proton-proton and proton-neutron interactions, respectively. The proton-hole states in the present work are affected by the missing correlations due to the removal of protons from the fully occupied $28$-$50$ shell. In the following discussion, the configurations are written in the proton-hole neutron-particle scheme for simplicity.

\section{\label{sec:level3}Results and Discussion}

\begin{table}
\begin{center}
\caption{\label{sn}One neutron separation energies of experiments $S_{n}^{(expt)}$ from Ref~\cite{audi2012} except $^{131}$Cd from \cite{At15}, predictions $S_{n}^{(AME2012)}$ of AME2012~\cite{audi2012}, calculations through the finite range liquid drop model $S_{n}^{(FRDM)}$~\cite{moller1995} and the present work $S_{n}^{(jj46)}$. (All values are in MeV.)}
\begin{tabular}{ccccc}
\hline
  Nuclei & $S_{n}^{(expt)}$ & $S_{n}^{(jj46)}$ & $S_{n}^{(AME2012)}$ & $S_{n}^{(FRDM)}$ \\
\hline
$^{133}$Sn    & 2.402(4)   &   2.408     &         &  2.651    \\
$^{134}$Sn    & 3.629(4)   &   3.732     &         &  4.281    \\
$^{135}$Sn    & 2.271(4)   &   2.405     &         &  1.871    \\
$^{136}$Sn    &            &   3.795     &  3.340  &  3.741    \\
$^{137}$Sn    &            &   2.339     &  1.960  &  1.611    \\
$^{138}$Sn    &            &   3.834     &  3.140  &  3.561    \\
$^{132}$In    &2.450(60)   &   2.364     &         &  2.701    \\
$^{133}$In    &            &   3.418     &  3.130  &  3.781    \\
$^{134}$In    &            &   2.370     &  2.270  &  1.771    \\
$^{131}$Cd    &2.169(103)  &   1.984     &  1.870  &  2.031    \\
$^{132}$Cd    &            &   3.324     &  3.000  &  3.671    \\
$^{133}$Cd    &            &   1.977     &  1.730  &  1.321    \\
$^{130}$Ag    &            &   1.902     &  1.780  &  2.001    \\
$^{129}$Pd    &            &   1.524     &         &  1.461    \\
\hline
\end{tabular}
\end{center}
\end{table}

With the $jj46$ Hamiltonian described above, the properties of $^{133-138}$Sn, $^{131-134}$In, $^{130-133}$Cd,  $^{129-131}$Ag, and $^{128-130}$Pd are investigated. One-neutron separation energies calculated using the present Hamiltonian are listed in Table \ref{sn} together with the predictions of AME2012~\cite{audi2012}, the finite range droplet model (FRDM)~\cite{moller1995} results, and experimental values available. The present calculations reproduce the few experimental neutron separation energies~\cite{audi2012} in this region nicely. Both the single-particle energy of the $\nu1f_{7/2}$ orbit and the proton-neutron interactions involving the fully occupied proton orbits contribute to $S_{n}^{(jj46)}$ of $^{133}$Sn. Its value together with the other observed $S_{n}$ values are used to constrain the strength of the proton-neutron interaction in the present Hamiltonian as discussed in the previous section.

\begin{figure}
\begin{center}
\includegraphics[scale=0.12]{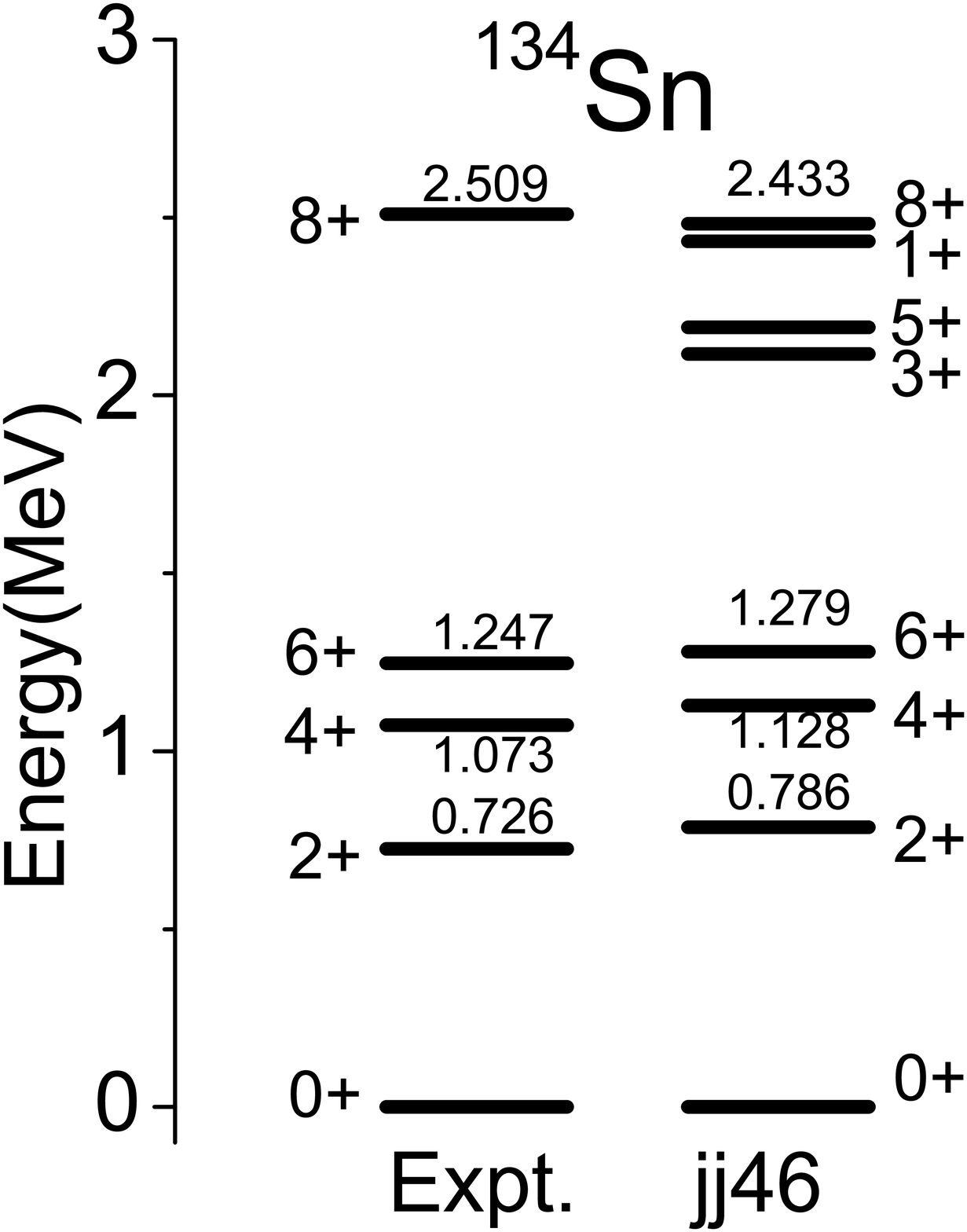}
\includegraphics[scale=0.12]{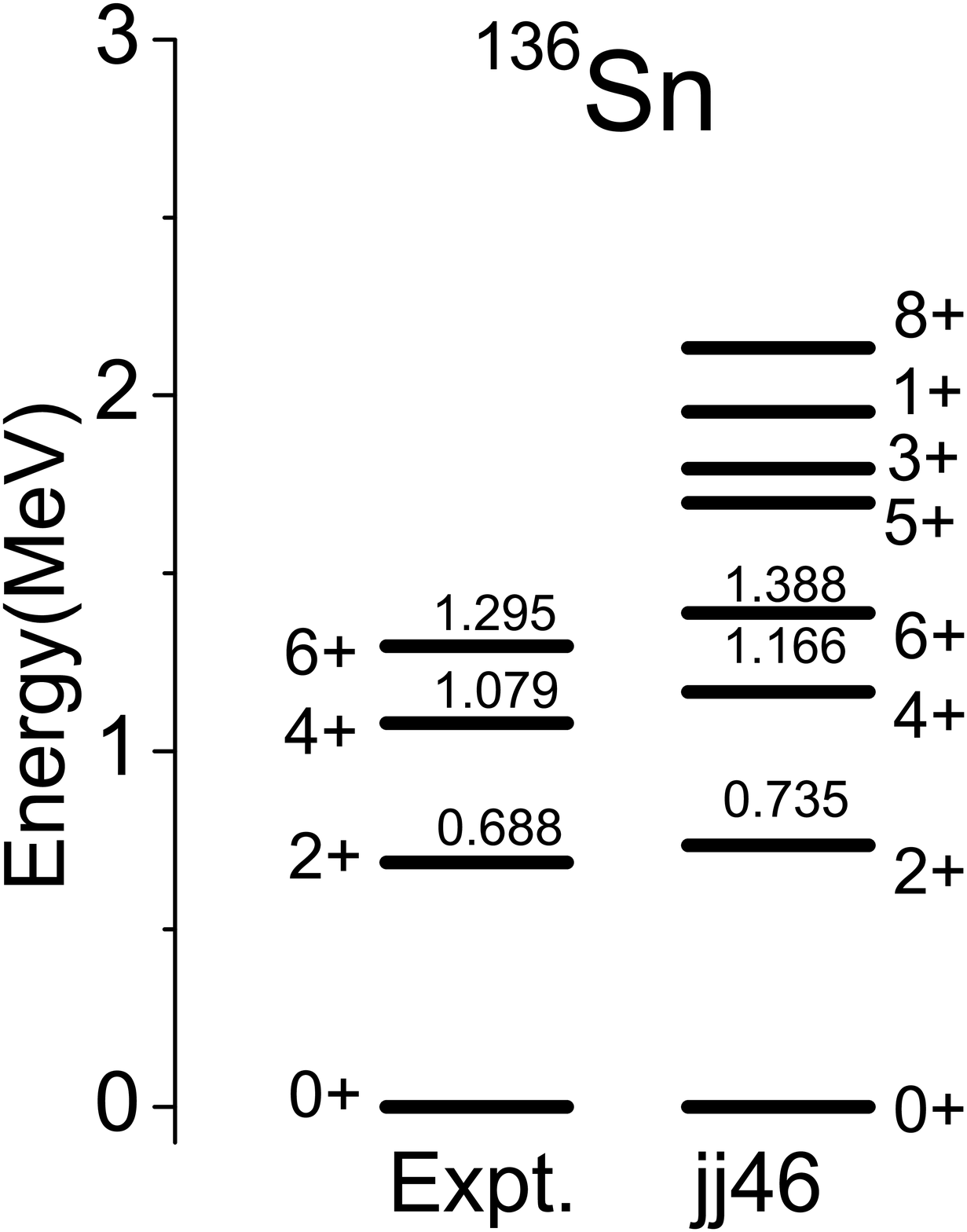}
\includegraphics[scale=0.12]{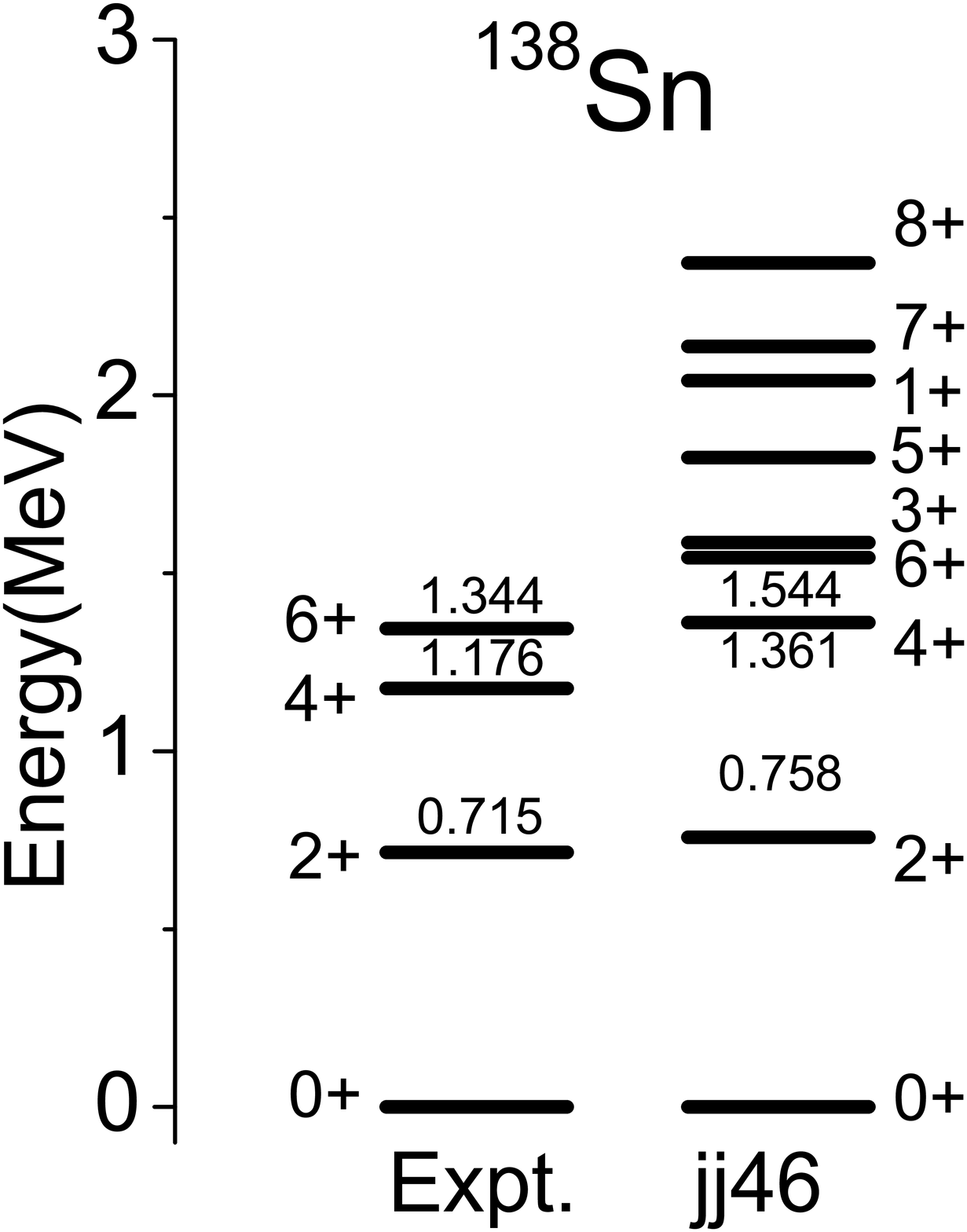}
\includegraphics[scale=0.12]{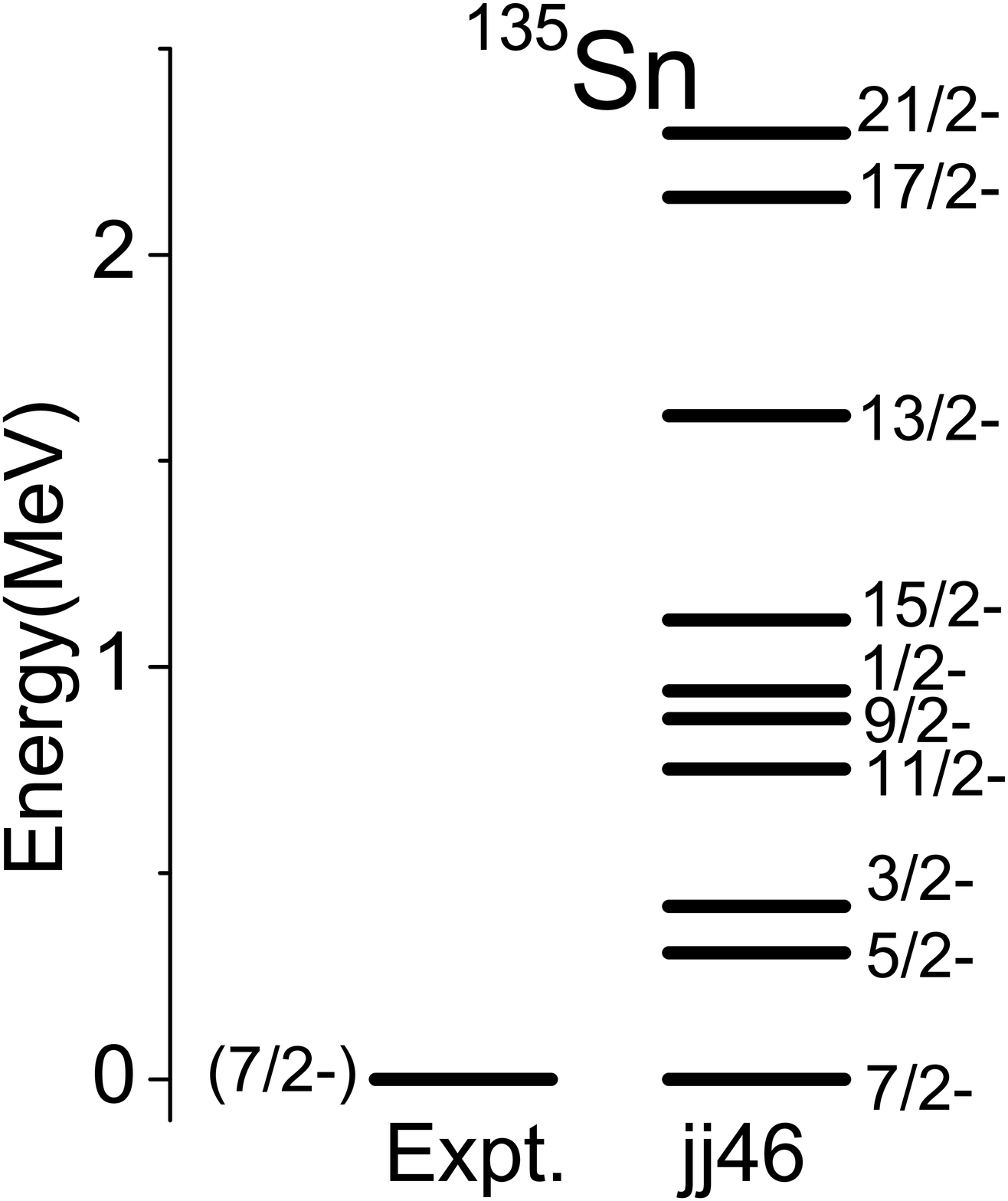}
\caption{\label{levels1} Comparison between the calculated levels in $^{134-136,138}$Sn in the present work and those observed experimentally~\cite{simpson2014,nndc}.}
\end{center}
\end{figure}

\begin{figure}
\begin{center}
\includegraphics[scale=0.24]{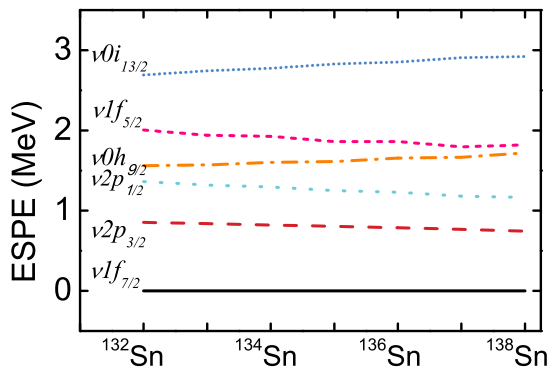}
\caption{\label{Z=50NESPE} Calculated neutron effective single-particle energies (ESPE) for Sn isotopes (Color online).}
\end{center}
\end{figure}

The levels of neutron-rich Sn isotopes are presented in Fig.~\ref{levels1}. The present Hamiltonian reproduces the known low-lying states of $^{134,136,138}$Sn well, especially the positions of the $2^{+}$ states in $^{134,136,138}$Sn and the increasing trend of the energies of $4^{+}$ and $6^{+}$ states from $^{134}$Sn to $^{138}$Sn. The dominant configurations of the $8^{+}$ states in $^{134}$Sn and $^{138}$Sn are 99.55\% $\nu(1f_{7/2})(1h_{9/2})$ and 41.83\% $\nu(1f_{7/2})^{5}(1h_{9/2})$, respectively. This state in $^{136}$Sn is dominated by the $\nu(1f_{7/2})^{4}$ configuration (84.34\%). If all the yrast $6^{+}$ and $4^{+}$ levels in $^{134,136,138}$Sn were dominated by a pure seniority-2 configuration, the $B(E2;6^{+}\rightarrow4^{+})$ value in $^{136}$Sn would be expected to be the lowest among these three isotopes, but their experimental results are decreasing from $^{134}$Sn to $^{138}$Sn. 
The seniority scheme of the low-lying states in these three isotopes can be discussed through the comparison between the experimental results and the shell model calculations. The observed $B(E2;6^{+}\rightarrow4^{+})$ value indicates a mixing of seniority-2 and -4 configurations in the $4^{+}$ state of $^{136}$Sn by comparing the results from a realistic effective interaction and the empirical modification of $\nu(1f_{7/2})^{2}$ matrix elements~\cite{simpson2014,Maheshwari2015}. The present calculation also gives the decreasing $B(E2)$ values between $6^{+}$ and $4^{+}$ states from $^{134}$Sn to $^{138}$Sn, which will be shown later.

The large energy differences between the $6^{+}$ and $8^{+}$ states in these three nuclei suggest that the $8^{+}$ states are not isomeric. However the small energy difference between the $17/2^{-}$ and $21/2^{-}$ states in $^{135}$Sn implies a $21/2^{-}$ metastable state. Details for this possible isomer in $^{135}$Sn will be given later. However, no such isomer is predicted in $^{137}$Sn (not shown in Fig.~\ref{levels1}). The first $21/2^{-}$ state of $^{135}$Sn is dominated by a $\nu(1f_{7/2})^{2}(0h_{9/2})$ configuration (96.7\%).
Compared with $^{135}$Sn the first $21/2^{-}$ level in $^{137}$Sn is expected to be more mixed because of the increasing number of valence neutrons and/or the change of the shell structure.
Fig.~\ref{Z=50NESPE} presents the effective single-particle energies (ESPE) of the neutron orbits in Sn isotopes. ESPE are defined as~\cite{otsuka20012},
\begin{equation}\label{speequ}
 \varepsilon_{j}=\varepsilon_{j}^{core}+\sum_{j'}V_{jj'}\langle\psi|\widehat{N}_{j'}|\psi\rangle,
\end{equation}
where $\varepsilon_{j}^{\rm core}$ is the single-particle energy relative to the core, $\langle\psi|\widehat{N}_{j'}|\psi\rangle$ is the shell-model occupancy of the $j'$ orbit and $V_{jj'}$ is the monopole part of the two-body interaction.
As shown in Fig.~\ref{Z=50NESPE}, the single particle energies do not change much in the Sn isotopes. Therefore the main difference between the $21/2^{-}$ states in $^{135}$Sn and $^{137}$Sn arises from the two additional valence neutrons in $^{137}$Sn. The configuration of the first $21/2^{-}$ level in $^{137}$Sn is a mixture of $\nu(1f_{7/2})^{4}(0h_{9/2})$ (61.3\%), $\nu(1f_{7/2})^{2}(2p_{3/2})^{2}(0h_{9/2})$ (9.97\%), $\nu(1f_{7/2})^{3}(2p_{3/2})(0h_{9/2})$ (9.05\%), and $\nu(1f_{7/2})^{2}(1f_{5/2})^{2}(0h_{9/2})$ (3.38\%), very different from the almost pure $\nu(1f_{7/2})^{2}\nu(0h_{9/2})$ configuration in $^{135}$Sn.

\begin{figure}
\begin{center}
\includegraphics[scale=0.12]{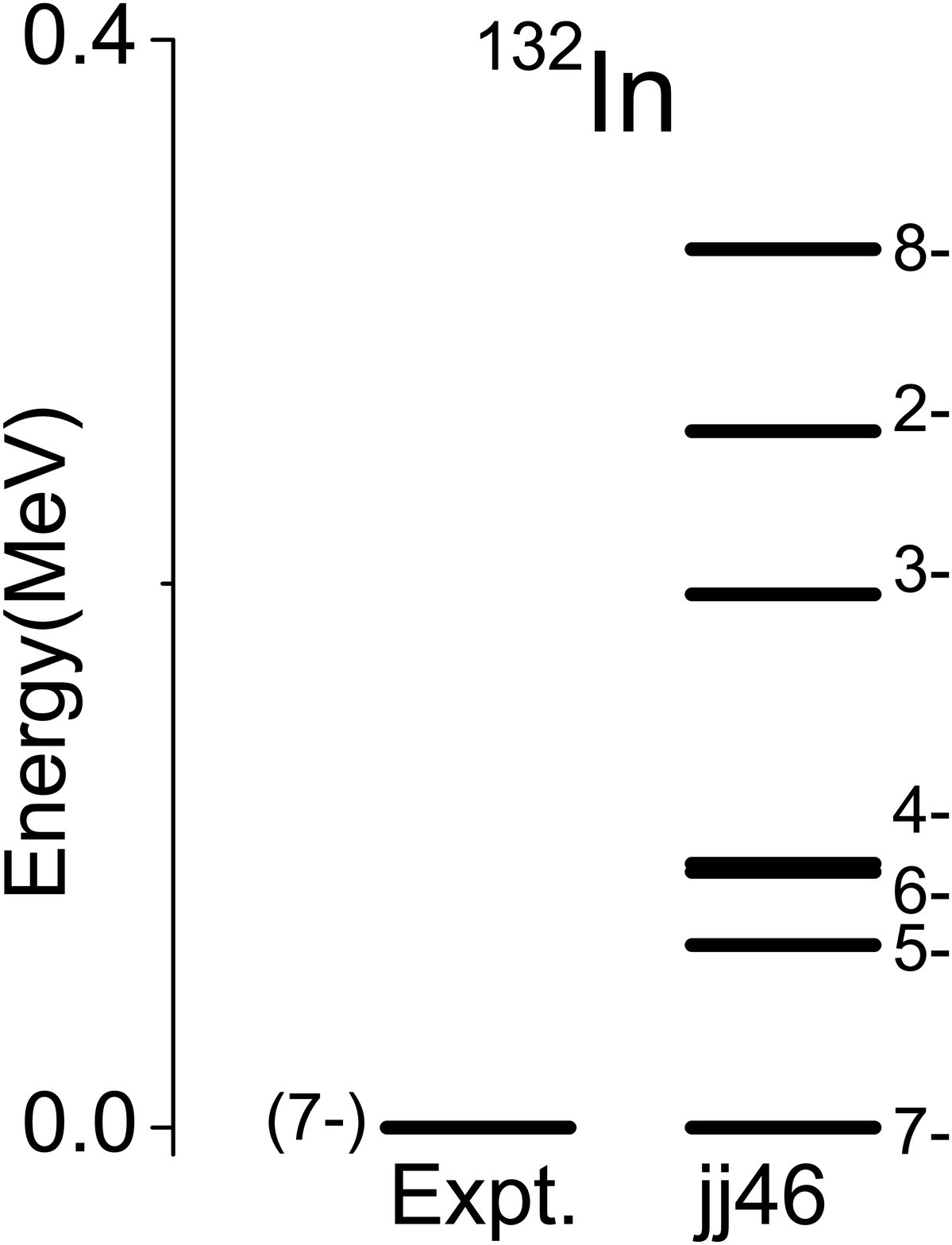}
\includegraphics[scale=0.12]{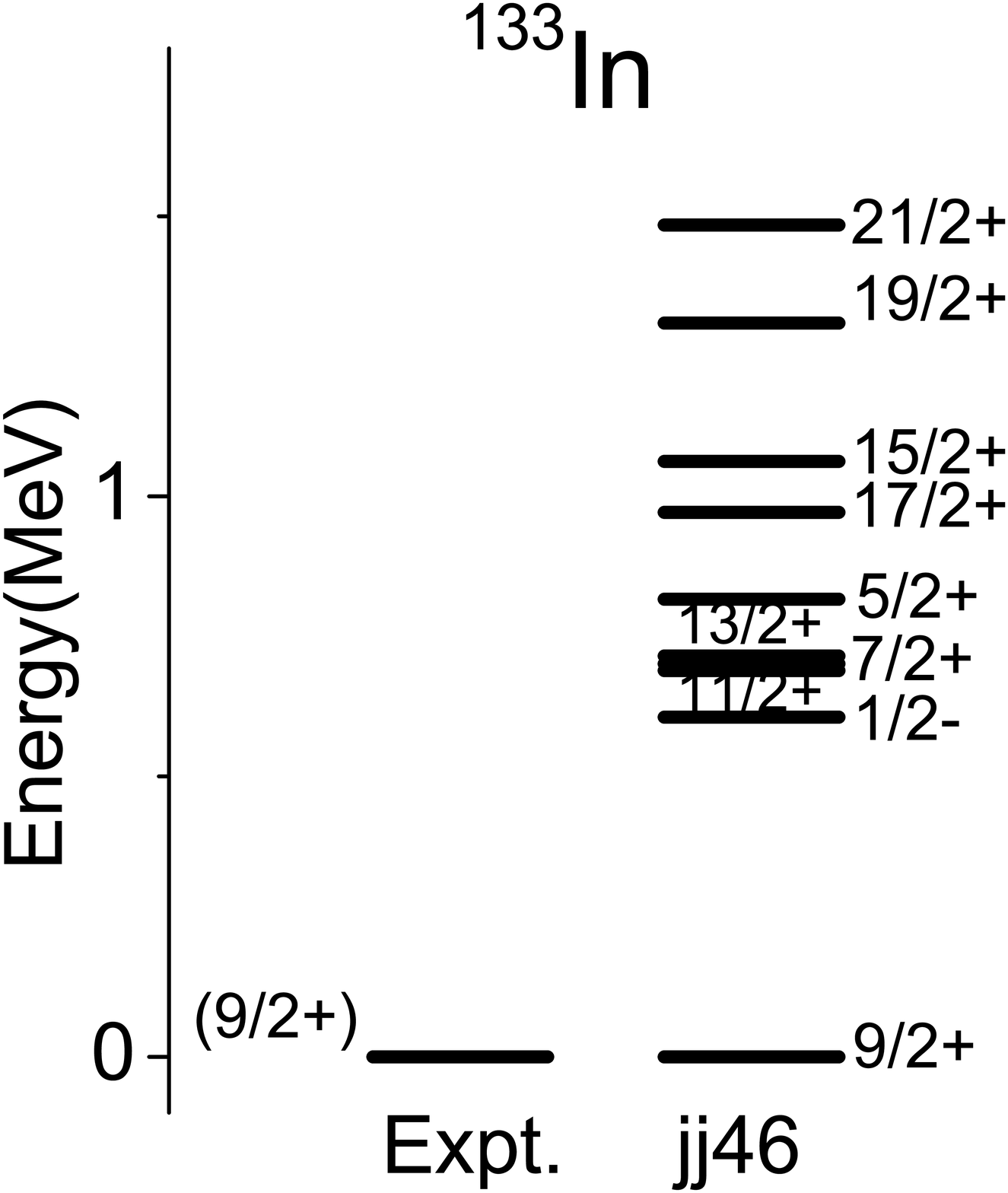}
\includegraphics[scale=0.12]{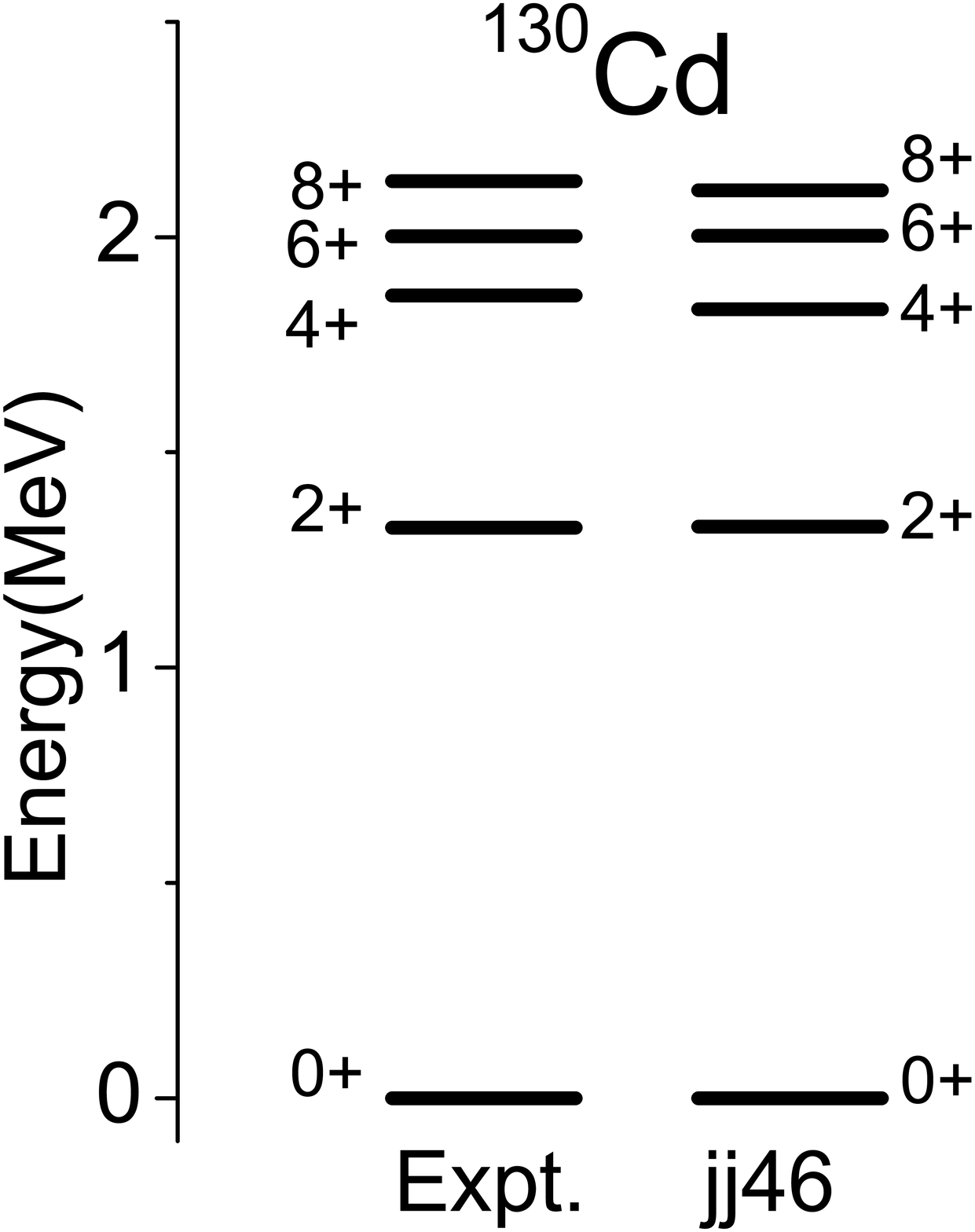}
\includegraphics[scale=0.12]{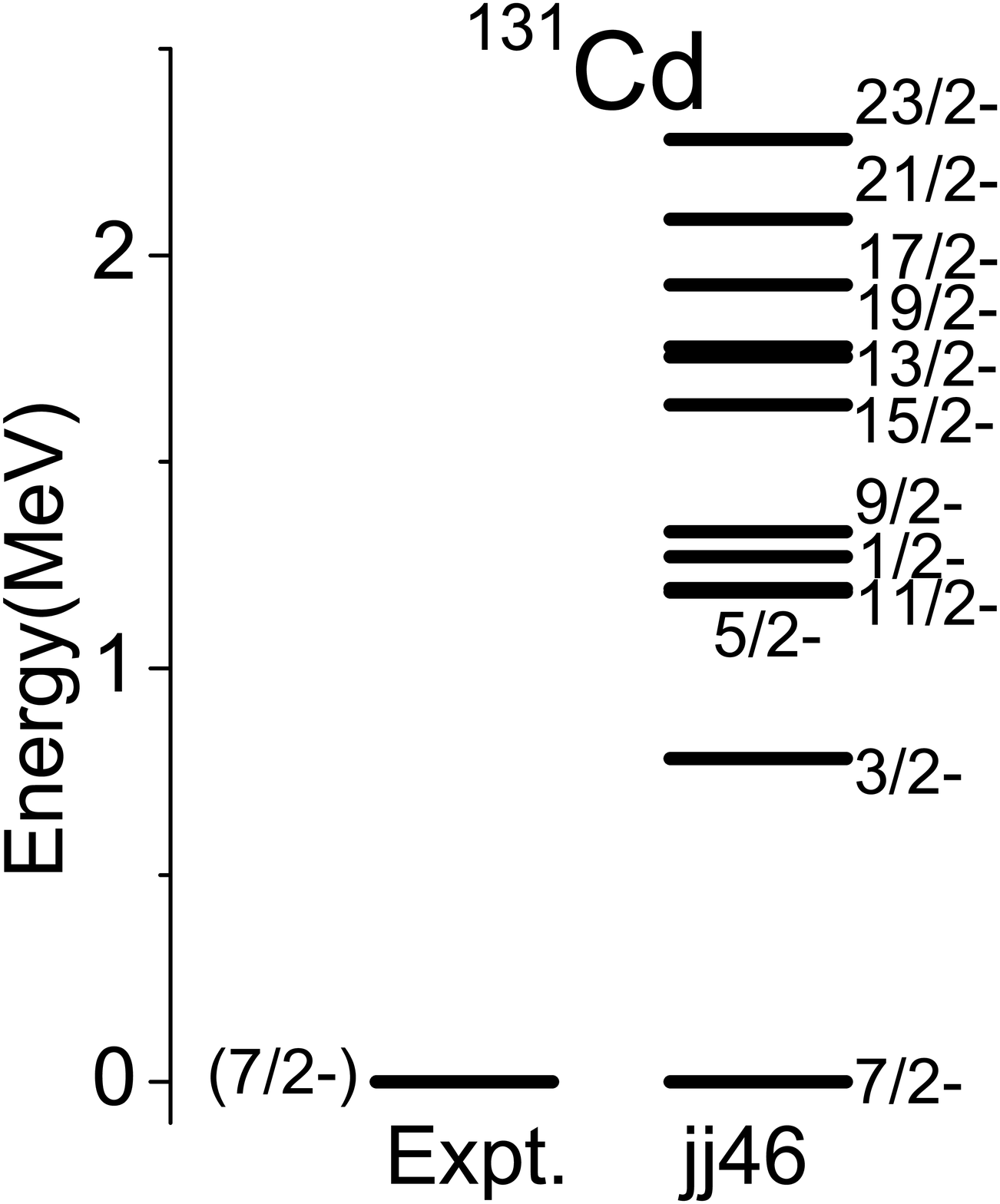}
\includegraphics[scale=0.12]{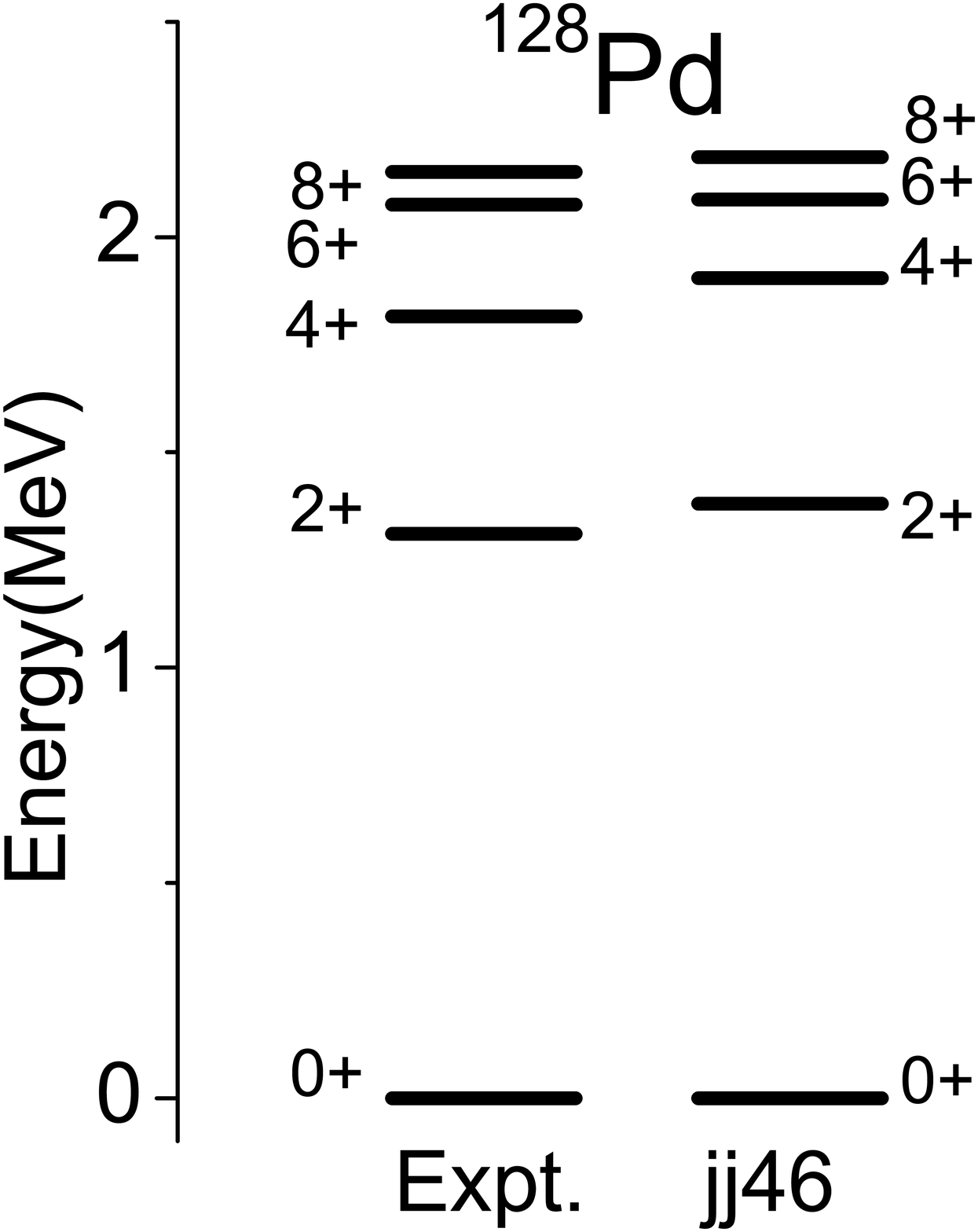}
\includegraphics[scale=0.12]{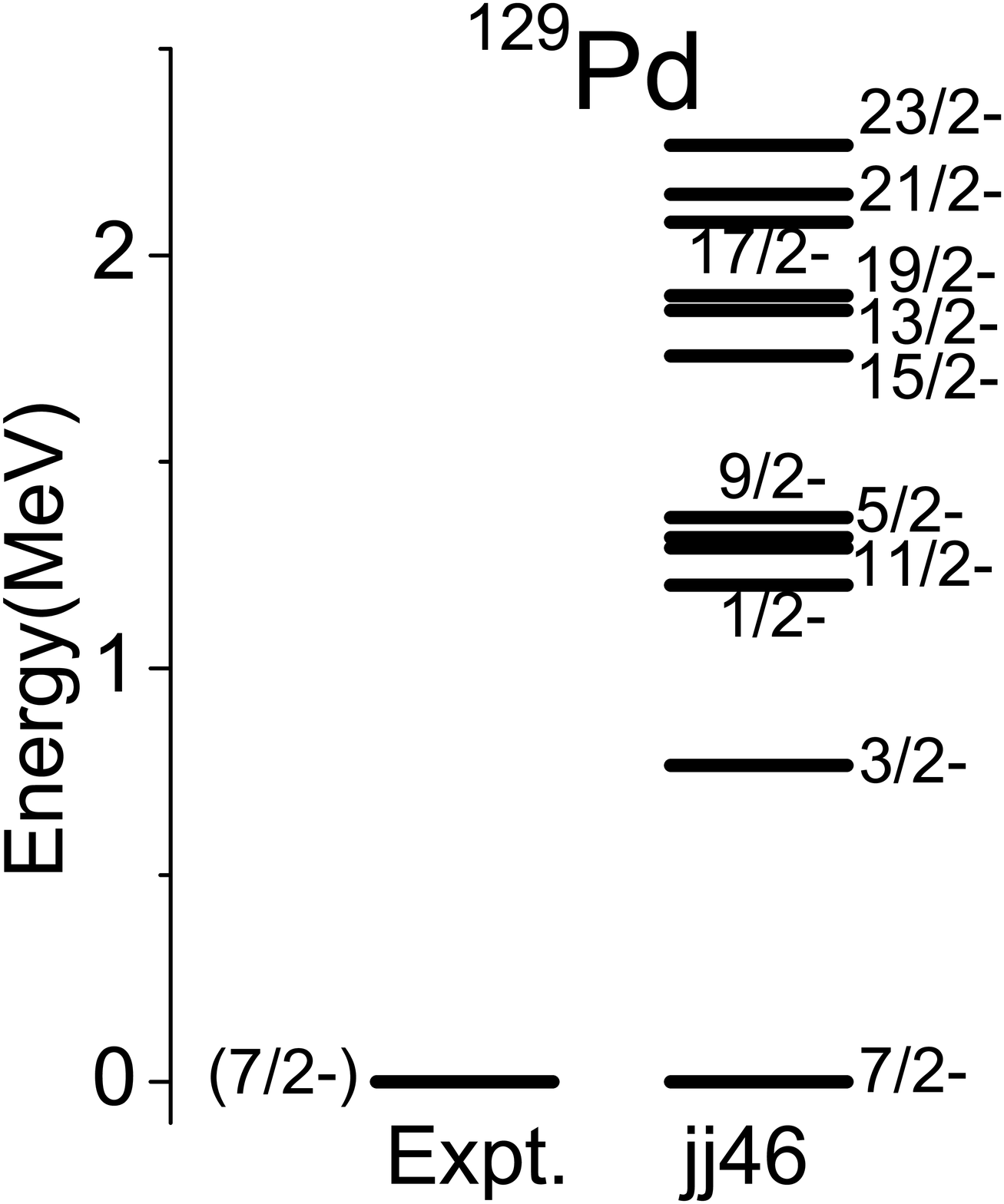}
\caption{\label{levels2} The same as Fig.~\ref{levels1}, but for $^{132,133}$In, $^{130,131}$Cd, and $^{128,129}$Pd.}
\end{center}
\end{figure}

\begin{figure}
\begin{center}
\includegraphics[scale=0.28]{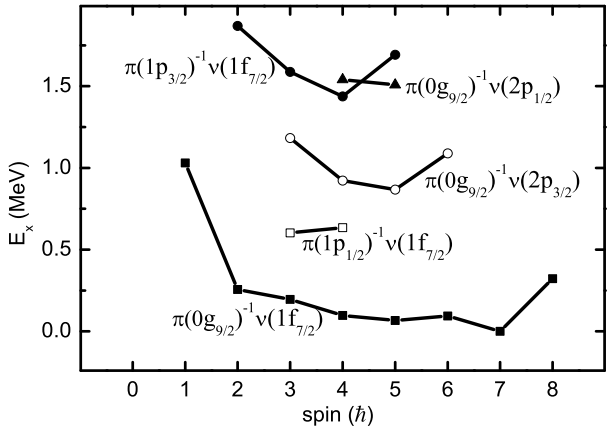}
\caption{\label{132In} Calculated excitation energies of the $\pi^{-1}\nu$ multiplets in $^{132}$In as function of spin.}
\end{center}
\end{figure}

\begin{figure}
\begin{center}
\includegraphics[scale=0.24]{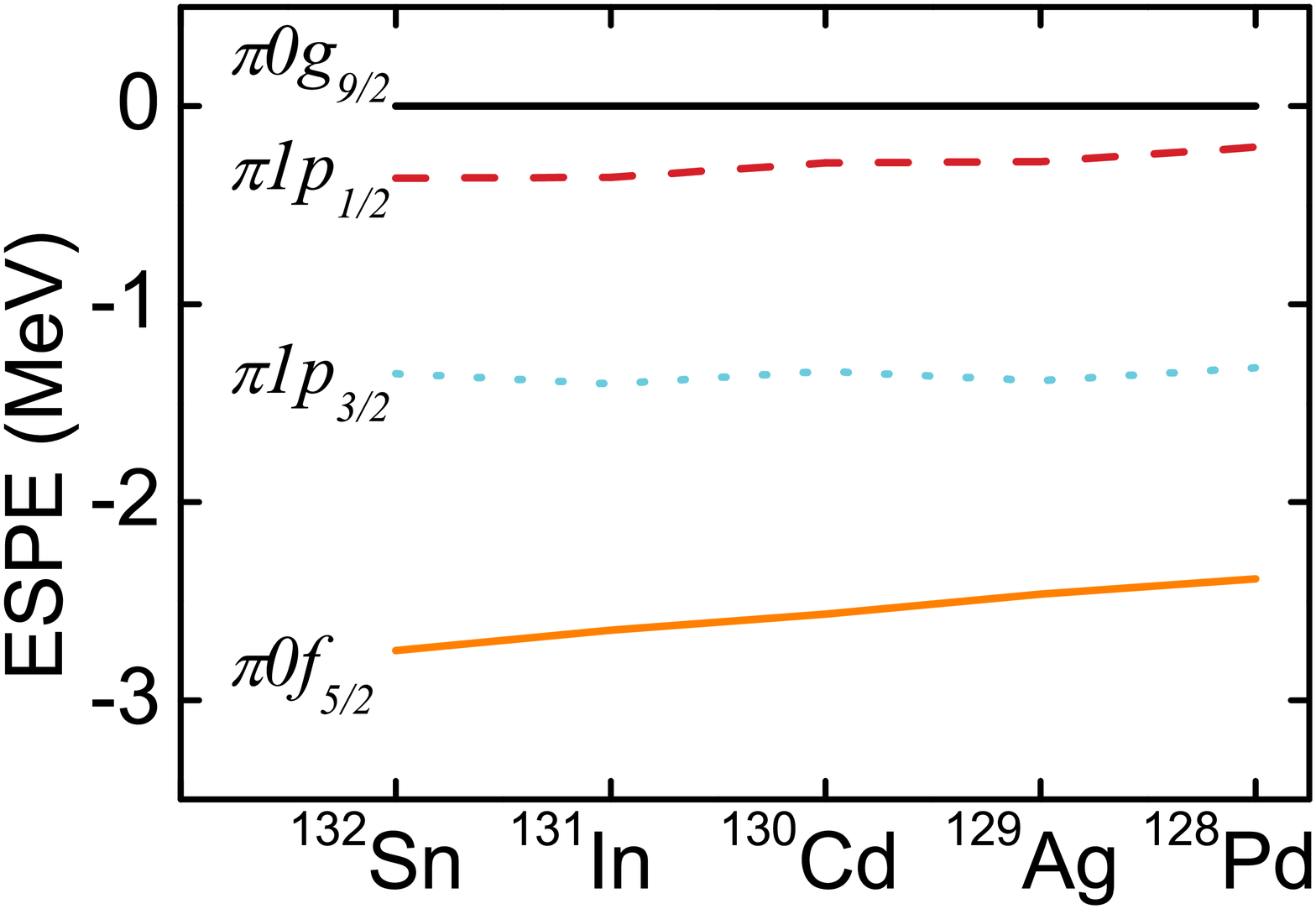}
\includegraphics[scale=0.24]{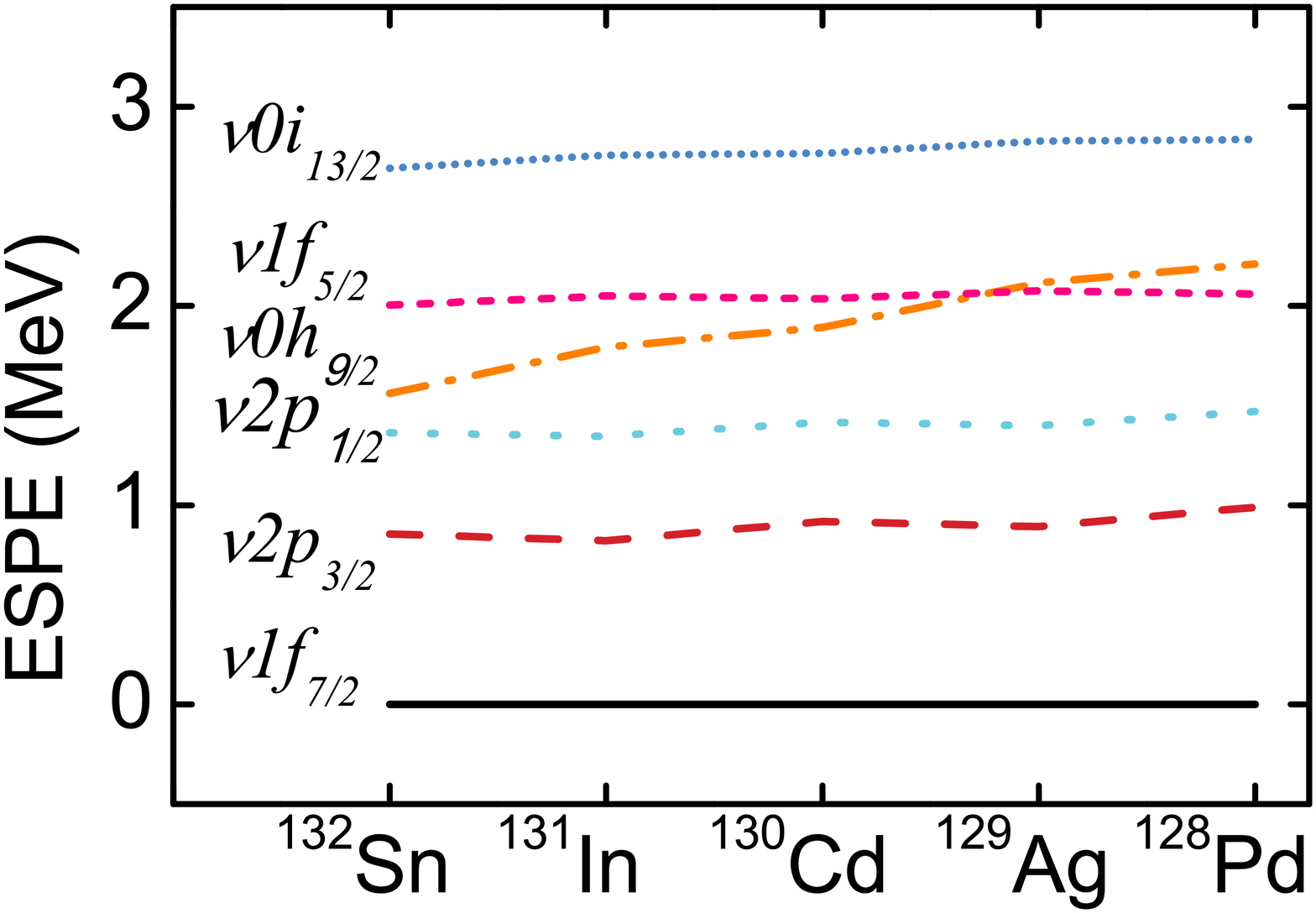}
\caption{\label{N=82ESPE} Proton hole and neutron particle ESPEs for $N=82$ isotones calculated in the present work (Color online).}
\end{center}
\end{figure}

Levels of the $^{132,133}$In, $^{130,131}$Cd, and $^{128,129}$Pd isotopes are presented in Fig.~\ref{levels2}. Some of them are possibly isomers.
The ground state of $^{132}$In is found to be $7^{-}$ with a configuration of $\pi(0g_{9/2})^{-1}\nu(1f_{7/2})$, in agreement with the experimental assignment~\cite{nndc}. Our calculations indicate a very low $5^{-}$ state in $^{132}$In, which can be a candidate for an isomer. With two more neutrons and two more proton holes respectively, $^{134}$In and $^{130}$Ag have similar structure, $7^{-}$ for ground state and very low $5^{-}$ for first excited state. These results depend on the details of the proton-neutron interaction between $\pi(0g_{9/2})^{-1}$ and $\nu(1f_{7/2})$ orbits, for which the experimental information is still rare.

Recently six $\gamma$ rays were observed following the $\beta$-delayed neutron emission from $^{133}$Cd and assumed to be emitted from the excited states of $^{132}$In~\cite{Jungclaus2016}. Due to the low statistics and the lack of $\gamma$-$\gamma$ coincidence, it was difficult to establish a level scheme with those observed $\gamma$ rays~\cite{Jungclaus2016}. The spin parity of the ground state of $^{133}$Cd is $7/2^{-}$ from Ref.~\cite{nndc} and the present shell-model calculation. As the $\beta$ decay energy ($\sim13$ MeV) of $^{133}$Cd is much larger than the neutron separation energy ($\sim3$ MeV) of $^{133}$In, the ground state of $^{133}$Cd can decay to many high excited states beyond the neutron emission threshold of $^{133}$In, with spin parity $5/2^{-}$ to $9/2^{-}$ through Gamow-Teller transitions and $3/2^{+}$ to $11/2^{+}$ through first forbidden transitions. Thus many states in $^{132}$In could be populated in the $\beta$-delayed neutron emission of $^{133}$Cd.

The low-lying states of the $\pi^{-1}\nu$ configurations in $^{132}$In calculated by the present Hamiltonian are shown in Fig.~\ref{132In}.
The present shell-model results are similar to those shown in Fig.~4(a) of Ref.~\cite{Jungclaus2016}, especially for the $\pi(1p_{1/2})^{-1}\nu(1f_{7/2})$, $\pi(0g_{9/2})^{-1}\nu(2p_{3/2})$, and $\pi(1p_{3/2})^{-1}\nu(1f_{7/2})$ multiplets, though the proton-neutron interaction is calculated through V$_{MU}$+LS in this work, while it is derived from the CD-Bonn potential with V$_{low-k}$ approach in \cite{Jungclaus2016}.


 In both calculations the $7^{-}$ state is the lowest and $1^{-}$ the highest among the $\pi(0g_{9/2})^{-1}\nu(1f_{7/2})$ multiplets, however the other states are slightly different. The present work predicts that $5^{-}$ is first excited state and a little lower than $6^{-}$ state, while the shell-model calculation in Ref.~\cite{Jungclaus2016} gave opposite prediction.
 Similarly the present shell-model results may also explain most of the observed $\gamma$ rays in $^{132}$In~\cite{Jungclaus2016}.
  Because of the uncertainty of the shell-model calculations and extra complexity introduced by the low-lying $\pi(1p_{1/2})^{-1}\nu(1f_{7/2})$ multiplets, more experimental information on the excited states of $^{132}$In, $^{134}$In and $^{130}$Ag are highly desired and crucial in understanding the proton-neutron interactions in this region.

The ESPE for proton-hole and neutron-particle orbits along the $N=82$ closed shell are shown in Fig.~\ref{N=82ESPE}, where it can be seen that the $\pi(0g_{9/2})$ and $\pi(1p_{1/2})$ proton orbits are quite close to each other. So the $1/2^{-}$ states in odd-Z N=82 isotones are predicted to be the first excited state beyond ground state $9/2^{+}$ and are long-lived isomers, consistently with the experimental observation that in both $^{131}$In and $^{129}$Ag the low-lying $1/2^{-}$ states are $\beta$-decaying isomers ~\cite{Kankainen2013,nndc}. The ground state of $^{130}$Cd is found to be a mixture of $\pi(0g_{9/2})^{-2}$ and $\pi(1p_{1/2})^{-2}$. The low-lying positive-parity states of $^{130}$Cd are formed through the coupling of $\pi(0g_{9/2})^{-2}$. The present calculations show that the $2^{+}$ and $4^{+}$ states of $^{130}$Cd are dominated by the $\pi(0g_{9/2})^{-2}$ configuration with 93.0\% and 99.2\%, respectively. The contribution from other configurations such as $\pi(1p_{1/2})^{-1}(1p_{3/2})^{-1}$, $\pi(1p_{1/2})^{-1}(0f_{5/2})^{-1}$, or $\pi(1p_{3/2})^{-2}$, is very small as they have much higher energies. In the present model space, the $6^{+}$ and $8^{+}$ states can only be formed from the $\pi(0g_{9/2})^{-2}$ configuration. Previous shell-model work showed similar results \cite{grawe2004} for $^{130}$Cd.
As expected from the seniority scheme, the yrast $8^{+}$ state in $^{130}$Cd was found to be isomeric~\cite{jungclaus2007}. Similarly the $8^{+}$ seniority isomer in $^{128}$Pd was recently identified at RIKEN~\cite{watanabe2013}.

\begin{table*}
\begin{center}
\caption{\label{isomer}Excitation energies (MeV), transition energies (MeV), observed~\cite{jungclaus2007,watanabe2013,simpson2014} and calculated $B(E2)$ values ($e^{2}fm^{4}$), calculated $E2$ mean life-time ($\mu s$), and the dominant configurations for possible isomeric states.}
\begin{tabular}{cccccccc}
\hline
  Nuclei & $J_{i}^{\pi}$ $\rightarrow$ $J_{f}^{\pi}$ &$E_{i}$& $\Delta E$ & $B(E2)_{th}$  & $\tau_{E2}$ & $B(E2)_{Expt}$  & Configuration\\
\hline
$^{134}$Sn&$   6^{+}$ $\rightarrow$ $   4^{+}$&1.267  &0.149    &41.72      &0.266       &35(6)         & $\nu(1f_{7/2})^{2}$(96.3\%)\\
$^{135}$Sn&$21/2^{-}$ $\rightarrow$ $17/2^{-}$&2.288  &0.147    &93.96      &0.127       &              & $\nu(1f_{7/2})^{2}(0h_{9/2})$(96.7\%)   \\
$^{136}$Sn&$   6^{+}$ $\rightarrow$ $   4^{+}$&1.388  &0.222    &15.43      &0.098       &24(4)         & $\nu(1f_{7/2})^{4}$(75.5\%)             \\
$^{138}$Sn&$   6^{+}$ $\rightarrow$ $   4^{+}$&1.544  &0.183    &12.80      &0.311       &19(4)         & $\nu(1f_{7/2})^{6}$(53.13\%)          \\
$^{132}$In&$   5^{-}$ $\rightarrow$ $   7^{-}$&0.067  &0.067    &1.75       &345.7       &              & $\pi(0g_{9/2})^{-1}\nu(1f_{7/2})$(99.0\%)    \\
$^{133}$In&$17/2^{+}$ $\rightarrow$ $13/2^{+}$&0.972  &0.257    &48.36      &0.015       &              & $\pi(0g_{9/2})^{-1}\nu(1f_{7/2})^{2}$(93.9\%)\\
$^{134}$In&$   5^{-}$ $\rightarrow$ $   7^{-}$&0.074  &0.074    &27.69      &13.28       &              & $\pi(0g_{9/2})^{-1}\nu(1f_{7/2})^{3}$(72.5\%)\\
$^{130}$Cd&$   8^{+}$ $\rightarrow$ $   6^{+}$&2.109  &0.106    &59.07      &1.032       &66(13)/50(10) & $\pi(0g_{9/2})^{-2}$(100.0\%)\\
$^{131}$Cd&$19/2^{-}$ $\rightarrow$ $15/2^{-}$&1.778  &0.140    &100.03     &0.152       &              & $\pi(0g_{9/2})^{-2}\nu(1f_{7/2})$(99.7\%)\\
$^{130}$Ag&$   5^{-}$ $\rightarrow$ $   7^{-}$&0.072  &0.072    &18.72      &22.53       &              & $\pi(0g_{9/2})^{-3}\nu(1f_{7/2})$(74.3\%)\\
$^{128}$Pd&$   8^{+}$ $\rightarrow$ $   6^{+}$&2.186  &0.099    &12.50      &6.866       &8.43(0.25)    & $\pi(0g_{9/2})^{-4}  $(71.1\%)\\
$^{129}$Pd&$19/2^{-}$ $\rightarrow$ $15/2^{-}$&1.903  &0.146    &12.88      &0.955       &              & $\pi(0g_{9/2})^{-4}\nu(1f_{7/2})$(74.2\%)\\
$^{130}$Pd&$   6^{+}$ $\rightarrow$ $   4^{+}$&1.328  &0.213    &185.41     &0.010       &              & $\pi(0g_{9/2})^{-4}\nu(1f_{7/2})^{2}$(54.6\%)\\
\hline
\end{tabular}
\end{center}
\end{table*}

All the seniority isomers experimentally observed in $^{134,136,138}$Sn, $^{130}$Cd and $^{128}$Pd are well reproduced. Their semi-magic nature validates the neutron-neutron and proton-proton parts of the shell-model interactions in the present work.
In some other nuclei in this region, some of the levels are possibly isomeric because of the slow transition rates resulting from low transition energies. These isomer candidates are listed in Table~\ref{isomer} together with those experimentally confirmed in Sn isotopes and N=82 isotones. $\tau_{E2}$ is the mean life-time of $E2$ transitions obtained from the theoretical reduced transition probability $B(E2)$ values and predicted transition energies. The experimental $B(E2)_{expt}$ values of $^{134,136,138}$Sn are from Ref.~\cite{simpson2014}. The two $B(E2)_{expt}$ values of $^{130}$Cd are due to the two possible decay transition energies~\cite{jungclaus2007}.
 The effective charges for calculation of $B(E2)$ values for protons and neutrons are $e_{p}=1.7e$ and $e_{n}=0.7e$, respectively, which are similar to those used in Ref.~\cite{simpson2014}.

The $21/2^{-}$ isomer in $^{135}$Sn is analogous to the $6^{+}$ isomer in $^{134}$Sn, but its $B(E2)$ value is much larger than that of $^{134}$Sn, as shown in Table.~\ref{isomer}. The one-body transition density of the neutron $1f_{7/2}$ orbit from $6^{+}$ to $4^{+}$ in $^{134}$Sn is almost the same as that from $21/2^{-}$ to $17/2^{-}$ in $^{135}$Sn, and the $B(E2)$ enhancement in $^{135}$Sn is mainly due to the transition from $0h_{9/2}$ to $1f_{5/2}$ and $1f_{7/2}$ to $1p_{3/2}$. Although the occupancies of $1f_{5/2}$ and $1p_{3/2}$ in the $17/2^{-}$ state are small, the large number of $0h_{9/2}$ and $1f_{7/2}$ particles in the $21/2^{-}$ state result in a large one-body transition density. The transition energy between $21/2^{-}$ and $17/2^{-}$ in $^{135}$Sn is predicted to be almost the same as that between $6^{+}$ and $4^{+}$ in $^{134}$Sn. $\tau_{E2}$ around $100$ ns is predicted for the $21/2^{-}$ isomer in $^{135}$Sn.

The yrast $5^{-}$ states in $^{132,134}$In and $^{130}$Ag are predicted to be closely below the $6^{-}$ levels and only around $70$ keV above the $7^{-}$ ground states, as shown in Fig.~\ref{132In} for $^{132}$In, leading to long life-times. A much larger $\tau_{E2}$ in $^{132}$In is predicted because of the small $B(E2)$ value caused by the cancellation between the proton and neutron contributions. It should be noted that due to the lack of experimental information on the proton-neutron interaction in this quadrant of $^{132}$Sn, large uncertainties related to the isomerism in these odd-odd nuclei are not unexpected in the present shell-model calculations. For example, the $5^{-}$ isomer will disappear if it is higher than the $6^{-}$ state. Or alternatively the $4^{-}$ state could be isomeric, if it lies below the $5^{-}$ state.

$19/2^{-}$ isomers are predicted in the $N=83$ isotones $^{131}$Cd and $^{129}$Pd, with $\tau_{E2}$ $\sim100$ ns and $\sim1$ $\mu$s, respectively. The excitation energy of the predicted $19/2^{-}$ isomer in $^{129}$Pd is around $0.4$ MeV above its neutron separation energies predicted by the present work and by Moller et al.~\cite{moller1995} (Table~\ref{sn}) and so it may decay to the ground state of $^{128}$Pd by emitting a neutron with an orbital angular momentum of $l$ = 9$\hbar$ (not to $2^{+}$ state of $^{128}$Pd because of its $1.311$ MeV excitation energy).
The half-life of neutron emission is calculated by using the widely-used formula of the two-potential approach~\cite{Gurvitz1987}, in which both the pre-exponential factor and exponential factor are explicitly defined. The potential that the emitted neutron feels is a sum of the nuclear potential, the spin-orbit potential and the centrifugl potential. The form of both the Woods-Saxon nuclear potential and the spin-orbit potential and the parameters used are taken from textbook~\cite{Nilsson1995},
\begin{equation}\label{potential}
 V=V_{central}(r)+\frac{\lambda}{r}\frac{\partial V_{central}(r)}{\partial r}\overrightarrow{l}\cdot\overrightarrow{s},
\end{equation}
where $V_{central}(r)=-\frac{V_{0}}{1+e^{\frac{r-R}{a}}}$, $R=1.2A^{1/3}$ fm, $a=0.5$ fm, $\lambda=-0.062\frac{2\hbar}{M\omega_{0}}$, $\hbar\omega_{0}=41A^{1/3}(1+\frac{1}{3}\frac{N-Z}{A})$, and $M$ is the average mass of a nucleon. The depth of the nuclear potential $V_{0}$ is determined by the Bohr-Sommerfeld condition to ensure a quasi-bound state \cite{Xu2006},
\begin{equation}\label{BScondition}
 \int^{r_{2}}_{r_{1}}\sqrt{\frac{2\mu}{\hbar^{2}}(Q_{n}-V-\frac{\hbar^{2}}{2\mu}\frac{l(l+1)}{r^{2}})}dr=(G-l+1)\frac{\pi}{2},
\end{equation}
where $\mu$ is the reduced mass of the neutron, $Q_{n}$ is the decay energy, $r_{1}$ and $r_{2}$ are classical turning points defined by $Q_{n}=V+\frac{\hbar^{2}}{2\mu}\frac{l(l+1)}{r^{2}}$, and $G$ is the global quantum number. The predicted life-time for neutron emission from the $19/2^{-}$ state of $^{129}$Pd to the ground state of $^{128}$Pd ranges from $0.3$ to $10$ $\mu$s with the decay energy $Q_{n}$ = 0.35-0.50 MeV, comparable to that of the calculated electromagnetic decay. The global quantum number in the Bohr-Sommerfeld condition is fixed at 9 due to the very low decay energy.


\section{\label{sec:level4}Summary}

In summary, a shell-model study has been performed in the south-east of $^{132}$Sn to search for possible isomeric states. A new shell-model Hamiltonian has been constructed for the present investigation. The proton-proton and neutron-neutron interactions, which are each limited to one major shell, have been obtained from existing CD-Bonn $G$ matrix calculations. The proton-neutron interaction across two major shells is calculated through the V$_{MU}$ plus M3Y spin-orbit interaction. The present Hamiltonian, $jj46$, is able to reproduce well the one-neutron separation energies, level energies, and $B(E2)$ values of the already observed isomers in this region. New isomeric states are predicted and their structures are discussed. The predicted $19/2^{-}$ isomer in $^{129}$Pd could be a candidate for neutron radioactivity.

\section{Acknowledgements}

This work has been supported by the National Natural Science Foundation of China under Grant Nos. 11235001, 11305272, 11375086, 11405224, 11435014, 11575007, 11535004, and 11320101004, the Special Program for Applied Research on Super Computation of the NSFC Guangdong Joint Fund (the second phase), the Specialized Research Fund for the Doctoral Program of Higher Education under Grant No.~20130171120014, the Guangdong Natural Science Foundation under Grant No.~2014A030313217, the Fundamental Research Funds for the Central Universities under Grant No.~14lgpy29, the Pearl River S\&T Nova Program of Guangzhou under Grant No.~201506010060 and Hundred-Talent Program (Chinese Academy of Sciences), and the United Kingdom Science and Technology Facilities Council under grant No. ST/L005743/1.




\begin{thebibliography}{00}




\bibitem{segre1949} E. Segre and A.C. Helmholz, Rev. Mod. Phys. {\bf 21} 271 (1949).
\bibitem{dillmann2003} I. Dillmann et al., Phys. Rev. Lett. {\bf 91}, 162503 (2003).
\bibitem{jungclaus2007} A. Jungclaus, \emph{et al.}, Phys. Rev. Lett. {\bf 99}, 132501 (2007).
\bibitem{gorska2009} M. G$\acute{o}$rska, \emph{et al.}, Phys. Lett. B {\bf 672}, 313 (2009).
\bibitem{watanabe2013} H. Watanabe, \emph{et al.}, Phys. Rev. Lett. {\bf 111}, 152501 (2013).
\bibitem{korgul2000} A. Korgul, \emph{et al.}, Eur. Phys. J. A {\bf 7}, 167 (2000).
\bibitem{simpson2014} G.S. Simpson, \emph{et al.}, Phys. Rev. Lett. {\bf 113}, 132502 (2014).
\bibitem{At15} D. Atanasov, \emph{et al.}, Phys. Rev. Lett. {\bf 115}, 232501 (2015).

\bibitem{Kn16} R. Kn\"obel, \emph{et al.}, Phys. Lett. B {\bf 754}, 288(2016).
\bibitem{jackson1970} K.P. Jackson \emph{et al.}, Phys. Lett. B {\bf 33}, 281 (1970).
\bibitem{Pe71} L.K. Peker, \emph{et al.}, Phys. Lett. B {\bf 36}, 547 (1971).
\bibitem{walker2006} P.M. Walker, AIP Conf. Proc. {\bf 819}, 16 (2006).


\bibitem{Jungclaus2016} A. Jungclaus, \emph{et al.}, Phys Rev. C {\bf 93}, 041301(R) (2016).

\bibitem{nndc} http://www.nndc.bnl.gov/nudat2/
\bibitem{Kankainen2013} A. Kankainen, \emph{et al.}, Phys Rev. C {\bf 87}, 024307 (2013).
\bibitem{taprogge2014} J. Taprogge, \emph{et al.}, Phys. Rev. Lett. {\bf 112}, 132501 (2014).
\bibitem{Urban1999} W. Urban, \emph{et al.}, Eur. Phys. J. A {\bf 5}, 239 (1999).

\bibitem{OXBASH}OXBASH, the Oxford, Buenos-Aires, Michigan State, Shell Model Program, B.A. Brown, A. Etchegoyan, and W.D.M. Rae, MSU Cyclotron
Laboratory Report No. 524, 1986.
\bibitem{Jensen1995} M. Hjorth-Jensen, T.T.S. Kuo, and E. Osnes, Phys. Rep. {\bf 261}, 125 (1995); A. Holt, T. Engeland, M. Hjorth-Jensen, and E. Osnes, Phys.
Rev. C {\bf 61}, 064318 (2000); Engeland, M. Hjorth-Jensen, and E. Osnes, Phys. Rev. C {\bf 61}, 021302(R) (2000).

\bibitem{Haaranen2016} M. Haaranen, P.C. Srivastava, and J. Suhonen, Phys Rev. C {\bf 93}, 034308 (2016).
\bibitem{Rejmund2016} M. Rejmund \emph{et al.}, Phys. Lett. B {\bf 753}, 86 (2016).
\bibitem{brown2005} B.A. Brown, N.J. Stone, J.R. Stone, I.S. Towner, and M. Hjorth-Jensen, Phys Rev. C {\bf 71}, 044317 (2005).

\bibitem{otsuka2010} T. Otsuka, T. Suzuki, M. Honma, Y. Utsuno, N. Tsunoda, K. Tsukiyama, and M. Hjorth-Jensen, Phys. Rev. Lett. {\bf 104}, 012501 (2010).
\bibitem{m3y1977} G. Bertsch, J. Borysowicz, H. McManus, and W. G. Love, Nucl. Phys. {\bf A284}, 399 (1977).
\bibitem{yuan2012} C.X. Yuan, T. Suzuki, T. Otsuka, F.R. Xu, and N. Tsunoda, Phys Rev. C {\bf 85}, 064324 (2012).
\bibitem{utsuno2012} Y. Utsuno, T. Otsuka, B. A. Brown, M. Honma, T. Mizusaki, and N. Shimizu, Phys Rev. C {\bf 86}, 051301(R) (2012).
\bibitem{Togashi2015} T. Togashi, N. Shimizu, Y. Utsuno, T. Otsuka, and M. Honma, Phys Rev. C {\bf 91}, 024320 (2015).
\bibitem{Utsuno2015} Y. Utsuno, N. Shimizu, T. Otsuka, T. Yoshida, and Y. Tsunoda, Phys. Rev. Lett. {\bf 114}, 032501 (2015).
\bibitem{Watanabe2014} H. Watanabe, \emph{et al.}, Phys. Rev. Lett. {\bf 113}, 042502 (2014).

\bibitem{audi2012} M. Wang, \emph{et al.}, Chin. Phys. C {\bf 36}(12), 1603 (2012).
\bibitem{otsuka20012} T. Otsuka, M. Honma, T. Mizusaki, N. Shimizu, and Y. Utsuno Prog. Part. Nucl. Phys. {\bf 47},319 (2001).

\bibitem{moller1995} P. M\"{o}ller, J.R. Nix, W. D. Myers, and W. J. Swiatecki, At. Data Nucl. Data Tables  {\bf 59}, 185 (1995).

\bibitem{Maheshwari2015} B. Maheshwari, A.K. Jain, and P.C. Srivastava, Phys Rev. C {\bf 91}, 024321 (2015).

\bibitem{grawe2004} H. Grawe, Lect. Notes Phys. {\bf 651}, 33 (2004).
\bibitem{Gurvitz1987} S.A. Gurvitz and G. Kalbermann, Phys. Rev. Lett. {\bf 59}, 262 (1987).


\bibitem{Nilsson1995} S.G. Nilsson and I. Ragnarsson, Shapes and Shells in Nuclear Structure (Cambridge University Press, 1995).
\bibitem{Xu2006} C. Xu and Z.Z. Ren, Phys. Rev. C {\bf 74}, 014304 (2006).















\end{thebibliography}


\end{document}